\documentclass[11pt]{article}
\pdfoutput=1

\usepackage{pstricks}
\usepackage{pst-plot}
\usepackage{pst-node}
\usepackage{amsmath}
\usepackage{pst-grad}
\usepackage{amssymb}
\usepackage{xspace}
\usepackage{pst-math}
\usepackage{pst-xkey}
\usepackage{pst-coil}
\usepackage{pst-3dplot}
\usepackage[T1]{fontenc}
\usepackage{ae}
\usepackage[ansinew]{inputenc}
\usepackage{graphics}
\usepackage{color}
\usepackage{float}
\usepackage{graphicx}

\newrgbcolor{darkred}{0.8 0 0}
\newrgbcolor{darkerred}{0.7 0 0}
\newrgbcolor{darkestred}{0.5 0 0}
\newrgbcolor{darkgreen}{0 0.5 0.5}
\newrgbcolor{darkergreen}{0 0.6 0}
\newrgbcolor{darkestgreen}{0 0.5 0}
\newrgbcolor{darkblue}{0 0 0.6}
\newrgbcolor{darkestblue}{0 0 0.5}
\newrgbcolor{redgreen}{0.8 0.8 0}
\newrgbcolor{bluegreen}{0 0.4 0.7}
\newrgbcolor{bluewhite}{0 0.9 1}
\newrgbcolor{lightgreen}{0.6 1 0.6}
\newrgbcolor{lightred}{1 0.8 0.8}
\newrgbcolor{lightyred}{1 0.4 0.4}
\newrgbcolor{lightyblue}{0.5 0.5 1}
\newrgbcolor{lightblue}{0.8 1 1}
\newrgbcolor{lilas}{0.5 0.3 0.7}
\newrgbcolor{lilaswhite}{1 0.8 1}
\newrgbcolor{redy}{1 0.6 0.6}
\newrgbcolor{bluish}{0.6 0.6 1}
\newrgbcolor{sand}{0.9 0.9 0.7}
\newrgbcolor{brown}{0.4 0.4 0}
\newrgbcolor{blue1}{0 0.35 0.65}
\newrgbcolor{blue2}{0 0.4 0.6}
\newrgbcolor{blue3}{0 0.45 0.55}
\newrgbcolor{blue4}{0 0.5 0.5}
\newrgbcolor{blue5}{0 0.55 0.45}
\newrgbcolor{blue6}{0 0.6 0.4}
\newrgbcolor{blue7}{0 0.65 0.35}

\begin{document}

\title{To lag or not to lag? How to compare indices of stock markets that operate at different times.}

\author{Leonidas Sandoval Junior \\ \\ Insper, Instituto de Ensino e Pesquisa}

\maketitle

\begin{abstract}
Financial markets worldwide do not have the same working hours. As a consequence, the study of correlation or causality between financial market indices becomes dependent on wether we should consider in computations of correlation matrices all indices in the same day or lagged indices. The answer this article proposes is that we should consider both. In this work, we use 79 indices of a diversity of stock markets across the world in order to study their correlation structure, and discover that representing in the same network original and lagged indices, we obtain a better understanding of how indices that operate at different hours relate to each other.
\end{abstract}

\section{Introduction}

Probably the major issue when dealing with stock markets around the globe is how to treat the differences in operating times of the stock exchanges. It is quite clear that financial markets in Pacific Asia often influence financial markets in the American continent, and that the New York Stock Exchange influences the next day of the Tokyo Stock Exchange, but how can one deal with these influences when, for example, one is trying to build a portfolio of international assets by minimizing the covariance matrix built from their time series? Which time series should be lagged with respect to others, if any?

Many authors studied the correlations between stock markets in the world, often using same-day or lagged correlations between financial market indices. In 1989, Eum and Shim \cite{ES1989} identified that innovations in the USA market transmit to other markets on the following day, and that the same markets on the same day exhibit low correlations. In 1990, Becker, Finnerty, and Gupta \cite{BFG1990} concluded that the correlation between the stock markets of Japan and of the USA increase when one considers the data from the USA lagged by one day, showing that there is more correlation between the stock market in the USA with the next day market of Japan. In 1994, Lin, Engle, and Ito \cite{LEI1994}, using intraday data that define daytime and overnight returns for the stock markets of the USA and Japan, found that the Tokyo daytime returns are correlated with the New York overnight returns. In 1996, Brailsford \cite{B1996} analyzed time-zone differences in trading hours between Australia and New Zealand in order to test for spillover effects, with the overnight returns from the USA market used to account for the impact of international news, showing that volatility surprises in the larger Australian market influence the subsequent conditional volatility of the smaller New Zealand market.

In 2000, Bonanno, Vandewalle, and Mantegna \cite{BVM2000} studied the effects of different currencies and differences in operation times on the correlations of stock markets; Mian and Adam \cite{MA2000} found that a larger proportion of the information affecting the Australian equities in fact arrives during the overnight closing period of the ASX, and discovered that the primary source of this anomalous finding is the dominant influence of the information originating in the USA; Vandewalle, Boveroux, and Brisbois \cite{VBB2000} studied the signs (up or down) of the indices Dow Jones (USA), DAX (Germany) and Nikkei (Japan), concluding that there is a domino effect in which changes in one stock market influence the next one, according to their opening hours.

In 2001, Dro\.{z}d\.{z}, Grümmer, Ruf, and Speth \cite{DGRS2001} studied a set of data with stocks that are part of the Dow Jones (USA) and of the DAX (Germany), and verified that, by properly taking into account the time-zone delays, both these markets largely merge into a single one, with the Dow Jones taking a leading role; Maslov \cite{M2001} made an analysis of daily returns of stock market indices in a diverse set of 37 countries worldwide, and observed that, for Asian stock markets, the correlation of their indices is larger when they are set against the American indices of the previous day. In 2006, Mayya and Amritkar \cite{MA2006} constructed and analyzed symmetrized delay correlation matrices for empirical data sets for atmospheric and financial data, finding that, for the financial time series, there is little correlation between different entities over a time delay beyond about two days; Kwapie\'{n}, Dro\.{z}d\.{z}, Górski, and O\'swi\c{e}cimka \cite{KDGO2006} found that the American and German stock markets evolve almost simultaneously without a significant time lag so that it is hard to find imprints of information transfer between the two markets.

In 2011, Song, Tumminello, Zhou, and Mantegna \cite{STZM2011} investigated the correlation of 57 stock market indices in periods of different lengths, and discovered that the (same-day) correlations among market indices presents both a slow dynamics (associated with the development and consolidation of globalization) and a fast one (associated with events that originate in specific parts of the world and that rapidly affect the global system). In 2012, Livan and Rebecchi \cite{LR2012} used daily prices of stocks belonging to the stock exchanges of the USA and of the UK, and looked for the emergence of correlations between the two markets in the eigenvalue spectrum of their non symmetric correlation matrix, also considering time-lagged correlations over short lags. In 2013, Nobi, Maeng, Ha, and Lee \cite{NMHL2013} used same-day cross-correlations in order to study some global and local financial market indices; Li \cite{L2013} found that asymmetric comovements between upturns and downturns of stock markets exist between the USA stock market and the stock markets of Canada, France, Germany, and the United Kingdom, but not necessarily between the USA stock market and the Japanese stock market.

Other methods for dealing with time series of data that is not synchronized were also developed and applied to international stock markets by other researchers. Lo and MacKinlay \cite{LM1990} developed an econometric analysis of nonsynchronous trading between markets. De Jong and Nijman \cite{JN1997} analyzed the lead-lag relationship between the S\&P 500 index and futures written on it. Münnix, Schäfer and Guhr \cite{MSG2010} developed a method to compensate statistical errors in the calculation of correlations on asynchronous time series, and applied it to high frequency financial data. Zhen and Yamasaki \cite{ZY2012} found that the peak of the cross-correlation function of volatility time series between the EUA (European Union allowances) or the WTI (West Texas Intermediate) and stock market indicators show a significant time shift (more than 20 days). Huth and Abergel \cite{HA2012} studied lead/lag relations in the CAC 40 index (France) using the Hayashi-Yashida cross-correlation estimator \cite{HY2005}, which deals with the issue of asynchronous trading. Goswamia, Ambikaa, Marwanb, and Kurthsb \cite{GAMK2012} use Correlation of Probability of Recurrence (CPR) \cite{CPR1}\cite{CPR2}, which was originally devised to quantify phase synchronization between non-phase-coherent and non-stationary time series, in order to study the correlations between nine stock market indices by using a measure of how similar were the recursion patterns (essentially, values of the indices that were similar to previous values of the same index) of those indices. Bastos and Caiado \cite{BC2010} used a similar technique, Recurrence Quantification Analysis (RQA), for studying 46 stock market indices in times of crises. Kumar and Deo \cite{KD2012} used Random Matrix Theory \cite{rmt1} and Multi-Fractal Detrended Fluctuation Analysis \cite{mfdfa} in order to study the correlation of 20 financial markets around the world without time lags.

We have also studied the subject of correlations among international stock exchange indices, using same-day correlations, applying it to the study of how correlations change in times of crises \cite{Leocorr}, and network representations of them \cite{Leotree}\cite{Leocluster}. Original and lagged correlations are compared in \cite{Leocluster}, showing there were not substantial alterations to a network of stock exchange indices built with original data or with some indices lagged by one day.

In this article, we use the correlations between the log-returns of 79 stock markets in the world in order to probe different approaches in dealing with differences in operation hours of stock exchanges, and propose a new approach in which one considers original and lagged indices as separate time series, resulting in an enlarged correlation matrix. Section 2 explains which data is being used and some of the methodology. Section 3 performs a comparison between the differences of opening hours of stock exchanges worldwide and analyzes the correlations of the log-returns of their indices. Section 4 analyzes the structure of the probability distributions of the eigenvalues obtained from the correlation matrices of the log-returns for two periods of time with different volatilities, and discusses the structures of the eigenvectors associated with the largest eigenvalues of those correlation matrices. Section 5 shows two ways to deal with the subject of different operation times: considering weekly data or removing the first two largest eigenvalues. Section 6 employs an enlarged correlation matrix consisting of original and lagged indices and analyzes its eigenvalue and eigenvector structure. Section 7 builds graphs of a network obtained from the enlarged correlation matrix and analyzes some centrality measures based on it. The conclusions are drawn in Section 8.

\section{Data and methodology}

We consider the time series of the daily closing prices of 79 benchmark indices of stock exchanges worldwide. The data span the period that goes from 2003 to 2012. The criteria used in the collection of data were availability and variety of indices, which belong to developed, emerging, or frontier markets. The indices and the countries to which they belong are displayed in Table 1.

\vskip 0.6 cm

\scriptsize

\begin{tabular}{|c|l|l||c|l|l|} \hline N & Index & Country & N & Index & Country \\ \hline 1 & S\&P 500 & United States of America & 41 & MICEX & Russia \\ 2 & S\&P/TSX Composite & Canada & 42 & ISE National 100 & Turkey \\ 3 & IPC & Mexico & 43 & KASE & Kazakhstan \\ 4 & Bermuda SX Index & Bermuda & 44 & Malta SX Index & Malta \\ 5 & Jamaica SX Market Index & Jamaica & 45 & Tel Aviv 25 & Israel \\ \hline 6 & BCT Corp Costa Rica & Costa Rica & 46 & Al Quds & Palestine \\ 7 & Bolsa de Panama General & Panama & 47 & ASE General Index & Jordan \\ 8 & Merval & Argentina & 48 & BLOM & Lebanon \\ 9 & Ibovespa & Brazil & 49 & TASI & Saudi Arabia \\ 10 & IPSA & Chile & 50 & MSM 30 & Oman \\ \hline 11 & IGBC & Colombia & 51 & DSM 20 & Qatar \\ 12 & IGBVL & Peru & 52 & ADX General Index & United Arab Emirates \\ 13 & IBC & Venezuela & 53 & Karachi 100 & Pakistan \\ 14 & FTSE 100 & United Kingdom & 54 & SENSEX 30 & India \\ 15 & ISEQ & Ireland & 55 & DSE General Index & Bangladesh \\ \hline 16 & CAC 40 & France & 56 & Colombo All-Share Index & Sri Lanka \\ 17 & DAX & Germany & 57 & Nikkei 25 & Japan \\ 18 & ATX & Austria & 58 & Hang Seng & Hong Kong \\ 19 & SMI & Switzerland & 59 & Shangai SE Composite & China \\ 20 & BEL 20 & Belgium & 60 & TAIEX & Taiwan \\ \hline 21 & AEX & Netherlands & 61 & KOSPI & South Korea \\ 22 & OMX Stockholm 30 & Sweden & 62 & MSE TOP 20 & Mongolia \\ 23 & OMX Copenhagen 20 & Denmark & 63 & Straits Times & Singapore \\ 24 & OMX Helsinki & Finland & 64 & Jakarta Composite Index & Indonesia \\ 25 & OBX & Norway & 65 & KLCI & Malaysia \\ \hline 26 & OMX Iceland All-Share Index & Iceland & 66 & SET & Thailand \\ 27 & MIB-30 & Italy & 67 & PSEI & Philippines \\ 28 & IBEX 35 & Spain & 68 & VN-Index & Vietnam \\ 29 & PSI 20 & Portugal & 69 & S\&P/ASX 200 & Australia \\ 30 & Athens SX General Index & Greece & 70 & NZX 50 & New Zealand \\ \hline 31 & WIG & Poland & 71 & CFG 25 & Morocco \\ 32 & PX 50 & Czech Republic & 72 & TUNINDEX & Tunisia \\ 33 & SAX & Slovakia & 73 & EGX 30 & Egypt \\ 34 & CROBEX & Croatia & 74 & Nigeria SX All Share Index & Nigeria \\ 35 & OMXT & Estonia & 75 & Gaborone & Botswana \\ \hline 36 & OMXR & Latvia & 76 & Ghana All Share Index & Ghana \\ 37 & Budapest SX Index & Hungary & 77 & NSE 20 & Kenya \\ 38 & SOFIX & Bulgaria & 78 & SEMDEX & Mauritius \\ 39 & BET & Romania & 79 & FTSE/JSE Africa All Share & South Africa \\ 40 & PFTS & Ukraine & & & \\ \hline
\end{tabular}

\normalsize

\vskip 0.3 cm

\noindent {\bf Table 1.} Indices of stock markets considered in this article and the countries they belong to. The data were collected from a Bloomberg terminal.

\vskip 0.3 cm

The log-returns of these indices, defined as
\begin{equation}
\label{logrets}
R_t=\ln (P_t)-\ln (P_{t-1})\ ,
\end{equation}
where $P_t$ is the closing price of the index at day $t$ and $P_{t-1}$ is the closing price of the same index at day $t-1$, are used in order to compute correlations between all log returns. There are many correlation measures available, being the Pearson correlation the most used one, but our choice was the Spearman rank correlation, since it is better suited to analyze nonlinear correlations.

One may then calculate the eigenvalues and eigenvectors of the resulting correlation matrices, and compare them with the results obtained by a correlation matrix resulting from shuffled data in order to pinpoint differences between them. The matrix constructed from these correlations may also be used in order to compute a distance measure between the indices, and this distance matrix may then be used to draw a map of the network of indices.

\section{Correlations and working hours}

In this section we shall study the correlations between world stock market indices according to their working hours. Figure 1 shows the opening and closing hours of the 79 stock exchanges considered in this article, not taking into account possible differences in intervals of the year due to daylight saving time. The dashed line marks the opening and closing hours of the New York Stock Exchange (NYSE), which shall be considered as our benchmark. One may see that many of the markets, mainly those of Western countries, have a sizeable or nonnegative intersection of working hours with those of the NYSE. Most markets located on the Eastern hemisphere, though, have working hours that do not overlap with those of the NYSE. How this difference affects the correlations between the stock market indices is analyzed now.

Considering data from the year 2003, as an example, we may calculate the correlations of all indices with the S\&P 500 on the same day and of all indices with the S\&P 500 of the previous day. We use the Spearman rank correlation, since it captures best nonlinear relationships, but it can be shown empirically \cite{pruning} that the Pearson correlation is as good as the Spearman one in what concerns our set of data.

By doing so, we may notice that some correlations between the S\&P 500 and the other indices actually increase when we compare those indices with the S\&P 500 lagged by one day. In contrast, the correlation with other indices drop when we consider the lagged index. A brief comparison of the differences between original and lagged correlations reveals that most indices which have positive results for lagged correlation are those of countries whose intersections with the opening hours of the NYSE are nonexistent. For most of the other indices whose stock exchanges operate partially or totally in the same hours as the NYSE, correlation drops when one considers lagged data.

Figure 2 shows the correlations of all indices with the S\&P 500 of the NYSE, all calculated on the same days. Figure 3 shows the correlations of the same indices with the S\&P 500 of the previous day. Correlations of the S\&P 500 with the other indices are stronger for indices that have overlapping working hours with the NYSE, and most indices that have no overlapping hours with the NYSE usually have larger correlations with the S\&P 500 of the previous day. Similar analysis comparing the correlations of the indices with the S\&P 500 of the next day shows negligible correlations, ranging from $-0.22$ to $0.18$, for indices of stock markets that have some overlapping with the NYSE. The indices that have stronger correlations with the lagged value of the S\&P 500 are those of Bermuda, Jamaica, Panama, Venezuela, and Iceland, all with overlapping hours with the NYSE, but which are small markets, not very connected to the other stock markets, and maybe slow to react to changes; Czech Republic, Slovakia, Estonia, Latvia, Bulgaria, Romania, Ukraine, Kazakhstan, Malta, European countries with very little or no overlapping with the working hours of the NYSE; and Lebanon, Oman, Qatar, United Arab Emirates, India, Sri Lanka, Japan, Hong Kong, China, Taiwan, South Korea, Singapore, Indonesia, Malaysia, Thailand, Philippines, Vietnam, Australia, and New Zealand, all of them in the Eastern hemisphere; Morocco, Tunisia, Egypt, Kenya, and Mauritius, none of them overlapping with the working hours of the NYSE.

Some other feature of figures 2 and 3 is that the correlation between the stock market indices with the S\&P 500 has been growing in the last decade. This growth is particularly visible from 2007 onwards. In the following section, we shall analyze the data in two periods of five years: one from 2003 to 2007, pre-subprime crisis, and another from 2008 to 2012, post-subprime crisis, and well into the credit crisis. This may also be seen if one calculates the cross-correlations between the S\&P 500 index at time $t$ and the other indices (including itself) at time $t+\tau $, where $\tau $ is a lag that may be positive or negative. For $\tau =1$, as an example, one considers the correlation of the S\&P 500 with another index quotation at the next day; if $\tau =-2$, one then considers the correlation of the S\&P 500 with the value of the other index two days before. The results are plotted in Figure 4, where one can see that most correlations are close to zero if lags of more than two days (positive or negative) are considered. For indices of countries with intersecting operating times with the NYSE, the largest correlations occur for a lag $\tau =0$; for European indices, the correlations for $\tau =1$ become larger as the index belongs to countries farther from the American continent; for Asian and Eastenr indices, in general, the largest correlations occur for $\tau =1$.

\section{Eigenvalue probability distribution}

\vskip 0.3 cm

Another way to analyze the influence of different opening hours in the correlation between the time series of international stock exchange indices is by considering the eigenvalues of the correlation matrix. In order to do this, we calculate the correlation matrix of the 79 indices here considered using data for the period 2003 to 2007 and for the period from 2008 to 2012, so as to compare two periods of time with very different volatilities. We then obtain a frequency distribution of the eigenvalues of such matrix. This distribution is quite different from one that may be obtained by randomly shuffling the original time series so as to preserve their averages and standard deviations, but destroy any simultaneous relation between each time series \cite{Leocluster}. The random distribution approaches the theoretical result obtained using Random Matrix Theory \cite{rmt1}. The two graphs of Figure 5 show histograms for the eigenvalues of the correlation matrices for 2003-2007 (left graph) and for 2008-2012 (right graph). Together with the graphs are plotted as solid lines the probability distributions of the eigenvalues for 10,000 simulations with randomized data for each block of observations. The results of the simulations are very similar to the theoretical Mar\v{c}enku-Pastur distributions \cite{rmt3}.

One may notice that the largest eigenvalue for each block of data is much larger than the maximum limit expected from a distribution obtained from random data. There are also other eigenvalues that are located outside the region defined by the Mar\v{c}enku-Pastur theoretical distribution, or by simulations with randomized data. This feature is enhanced in Figure 6, where we plot the eigenvalues as vertical lines and the gray areas as the regions associated with noise. On the left, we have the eigenvalues for data collected in 2003-2007, and on the right, the eigenvalues for data collected in 2008-2012. Note that the largest eigenvalue for 2008-2012 is rather larger than the largest eigenvalue for 2003-2007, an effect of the higher volatility of this period. Also, for 2008-2012, there are three other eigenvalues that clearly detach themselves from the bulk, while for 2003-2007, there are two of them. Besides the eigenvalues that are above the values predicted for a random collection of data, there are many eigenvalues bellow the same prediction. This happens in both periods being analyzed, and is linked to the possibility of combining pairs of stock indices that are very correlated in order to build ``portfolios'' of very low risk \cite{LeoBovespa}.

More information on the meaning of the largest eigenvalues may be obtained if one plots the eigenvectors associated with them. Figure 7 shows the components of the eigenvectors, plotted as column bars, associated with the three largest eigenvalues for the data collected in 2003-2007. Eigenvalue $e_1$, associated with the largest eigenvalue, has a very distinct structure, as nearly all indices appear with positive values. The exceptions are indices from stock markets that are very small in terms of number of stocks and of volume of negotiations. This eigenvector is often associated with a {\sl market mode}, and a portfolio built by taking the value of each index as its weight would follow very closely the general movements of the international stock market. In fact, when compared with an index of the global stock market, like the MSCI World Index, the portfolio built in terms of eigenvector $e_1$ has Pearson correlation $0.77$.

Eigenvector $e_2$, related with the second largest eigenvalue, has a structure that is typical of stock markets that do not operate at the same time \cite{Leocluster}. It generally shows positive values for Western countries and negative values for Eastern ones, defining two basic blocks. Eigenvector $e_3$, associated with the third largest eigenvalue, shows strong positive peaks in North America and South America, and negative peaks in Europe. Eigenvalue $e_4$, associated with the fourth largest eigenvalue, has strong positive peaks in Arab countries in the Middle East and North Africa, and smaller negative ones for Pacific Asian countries. Other eigenvectors also show some structure, but it is quickly lost as their associated eigenvalues approach the region where noise dominates.

Figure 8 shows similar results for the eigenvectors associated with the four largest eigenvalues of the correlation matrix obtained from the log-returns of data in 2008-2012.
Eigenvector $e_1$ still represents a market mode, with all significant indices appearing with positive values, and eigenvector $e_2$ also shows two blocks, a Western and an Eastern one. Eigenvector $e_3$ shows a block formed by North and South American indices, a block of European indices, probably joined by the African indices that operate at the same hours as the Central European markets, and a third block, of Pacific Asian indices, and it is probably connected with a fine tuning of the difference in operation hours of markets. Eigenvector $e_4$ separates indices from the Americas, from Europe, Arab countries, and Pacific Asian ones.

Comparing the eigenvectors in figures 7 and 8, one may also notice that there were no substantial changes to the structure shown by them from the period 2003-2007 to the period 2008-2012, showing that there is some stability on the world stock market structure, even during periods of crises.

\section{Two ways of dealing with different operating hours}

One way of dealing with markets that operate at different times that is used by many researchers is to consider, instead of daily data, the average of the time series over a week. By doing so, it is expected that the lag effect may be eliminated. In order to verify if this is a valid strategy, we take averages of our log-returns over each week, and then calculate the correlation matrix between all stock market indices, their eigenvalues and eigenvectors. This led to 503 weeks along the ten years of data, from 2003 to 2012, for the 79 stock market indices. The results are represented by Figure 9 and Figure 10, where the eigenvalues of the resulting correlation matrix, and the eigenvectors corresponding to the four largest eigenvalues are plotted.

There still is a presence of an eigenvalue that is far larger than the others, and well outside the region considered as noise. A second largest eigenvalue is also present, and a third largest one is very close to the border of the region assigned to noise. The eigenvectors of the first four largest eigenvalues also show some sort of structure. The eigenvector associated with the highest eigenvalue, $e_1$, shows a structure where most indices appear with positive signs, and is again associated with a market factor. The eigenvector associated with the second highest eigenvalue, $e_2$, shows as negative signed the indices of North America, of some South American countries, of Western Europe, and of South Africa, what may be a residual of different hours of operation or may represent genuine blocks of influence, since we should have removed those differences by considering weekly data. Eigenvector $e_3$, associated with the third largest eigenvector, which borders the region dominated by noise, highlights indices from South America and from Pacific Asia, and eigenvector $e_4$, associated with the fourth largest eigenvector, which is already within the region considered as noise, basically highlights indices from the Middle East and some indices from Pacific Asia.

By using weekly averages of log-returns, we tried to eliminate the effect of the difference in operating hours of the international stock exchanges, but with inconclusive results. Another way to try to eliminate the dependence on time differences is by using regressions. It is usual to subtract the market mode of a series of data using the eigenvector associated with the largest eigenvalue in order to build a portfolio of assets (indices, in the case of this article), and removing the contributions of the resulting time series of the portfolio from the other time series by using the linear regression
\begin{equation}
\label{regression}
R_t=a+bI_t+E_t\ ,
\end{equation}
where $R_t$ are the log-returns defined in (\ref{logrets}), $I_t$ is the log-return of the index built with the eigenvector associated with the largest eigenvalue, and $E_t$ are the residues of the regression. By doing this, we obtain new time series based on the residues of this regression. From those time series, we may build a new correlation matrix and calculate a new set of eigenvalues and eigenvectors associated with it. As shown in Figure 11 (left), the new eigenvalues do not reach a very high value, like in the case of the original data. The eigenvectors for the two largest eigenvalues, as shown in Figure 12 (top two graphs), reveals that now the largest eigenvalue represents the different operating times between Western and Eastern countries, like the second highest eigenvalue did for the original data. The eigenvector associated with the second largest eigenvalue shows a distinction between indices of the Americas and indices of the rest of the world.

We may now remove the largest eigenvalue of the residues by making another regression like in (\ref{regression}), again with the eigenvector corresponding to the largest eigenvalue for the residues. By doing that, we expect to remove the West-East effect of different operating hours of the stock exchanges. Figure 11 (right) shows that the eigenvalue frequency distribution has now a maximum eigenvalue that is not so large when compared with the expected values for a randomized distribution. The bottom two graphs of Figure 12 show the eigenvectors associated with the two largest eigenvalues after the two largest eigenvalues of the original data are removed. They show clusters of different continents, apparently without the interference of the West-East effect.

The correlation matrix may also be used in order to analyze the internal structure of the network formed by the indices. Figure 13 (left map) shows a false color map of the correlation matrix for the log-returns of data collected from 2003 to 2012. Brighter shades of gray indicate larger correlation that occurs, as expected, between the indices and themselves. Darker shades indicate lower correlation and even anti-correlation in some cases. There are some clear brighter regions, the main one concerning Western and Central European indices, and another one related with the correlations among North American indices. There is also a region of regular correlation among Pacific Asian markets. So, correlation between indices is deeply based on geographical region \cite{Leocluster}. Other long bright areas indicate correlations between continents. The darker areas correspond to the indices of Central America and the Caribbean Islands, and of Arab and African countries.

The correlation matrix of the residues of the log-returns after the largest eigenvalue mode is removed in represented in a false color diagram, also in Figure 13 (center map). One can see a reduced cluster of Western and Central European indices, and clusters of American and of Pacific Asian indices. Also slightly delineated is now a cluster of indices of Arab countries. Few clusters survive the removal of the two largest eigenvalue modes, as can be seen from Figure 13 (right map). Only the correlations among some core indices from the Americas, Europe, and Pacific Asia are visible. The Arab cluster is also visible in this map.

The two ways that we have seen here for diminishing the effects of different operating hours between stock exchanges around the world may be useful in order to understand how stock markets relate with one another without that influence, but they do not offer us a view of the way the markets correlate, taking into account those differences in operation times. An alternative that is proposed in this article is to consider not just the stock market indices on the same day, but also their lagged versions by one day, as part of a larger network. This is done in the next two sections.

\section{Enlarged correlation matrix}

Lagging some indices may give better results when analyzing the structure of stock market correlations, but it is not optimal. A new alternative is proposed in this article, which is to consider the time series of the international market indices and the time series of the same indices, but lagged by one day. So, instead of 79 indices, we now have 158 of them, counting the original and the lagged ones. Since we have shown in Section 4 that there are not many relevant differences between the two blocks, pre and post subprime crisis, and since we are now dealing with more time series, we here consider the data from the whole period, 2003-2012, and we work now with the usual Pearson correlation, since the differences between its results and the ones of the Spearman rank correlation are negligible, and it is much faster to calculate.

Building a correlation matrix based on the 158 resulting time series, one obtains very interesting results. First, we plot in Figure 14 a histogram of the eigenvalues associated with the correlation matrix for the whole data, together with the curve corresponding to the frequency distribution of the eigenvalues of 10,000 simulations with randomized data. Figure 15 shows the eigenvalues of the correlation matrix, with the shaded region representing the possible values spanned by the simulations with randomized data.

The highest eigenvalue is still much detached from the region considered as noise, and the second highest one is also very large. Two more eigenvalues are also well outside the noise region, and three others are slightly detached from it. Figure 16 shows the elements of the eigenvectors associated with the four highest eigenvalues of the enlarged correlation matrix, with gray bars meaning negative results and white bars meaning positive values. Eigenvector $e_1$, associated with the largest eigenvalue, again corresponds to a market mode, with most components, with the exception of the very small stock markets, appearing with relatively similar positive values, with larger values corresponding to lagged indices. Eigenvector $e_2$, corresponding to the second largest eigenvalue, exhibits a structure that clearly outlines the original and the lagged indices. Eigenvector $e_3$, corresponding to the third largest eigenvalue, shows a region at the center consisting of lagged American and Western European indices with the same sign as Asian and Eastern European indices of the next day, separated from American and Western European indices and lagged Asian and Eastern European indices. Eigenvector $e_4$ delineates an American and Western European block (on the left), and a lagged Western European block (on the right).

\section{Network structure of original and lagged indices}

The correlation matrix may be used in order to produce networks where nodes are the objects related by it, indices of international stock exchanges, in our case. The graphs of such networks may be used in order to filter some of the information of the correlation matrix, and also some of the noise contained in it. One of the most common filtering procedures is to represent those relations using a {\sl Minimum Spanning Tree} (MST) \cite{mst01}-\cite{mst14}\cite{pruning}, which is a graph containing all indices, connected by at least one edge, so that the sum of the edges is minimum, and which presents no loops. Another type of representation is that of a {\sl Planar Maximally Filtered Graph} (PMFG) \cite{pmfg01}-\cite{pmfg08}, which admits loops but must be representable in two-dimensional graphs without crossings.

Yet another type of representation is obtained by establishing a number which defines how many connections (edges) are to be represented in a graph of the correlations between nodes. There is no limitation with respect to the crossing of edges or to the formation of loops, and if the number is high enough, then one has a graph where all nodes are connected to one another. These are usually called {\sl Asset Trees}, or {\sl Asset Graphs} \cite{asset01}-\cite{asset08}, since they are not trees in the network sense. Another way to build asset graphs is to establish a value (threshold) such that distances above it are not considered. This eliminates connections (edges) as well as indices (nodes), but also makes the diagrams more understandable by filtering both information and noise. Some previous works using graphic representations of correlations between international assets (indices or otherwise) can be found in \cite{mst05}, \cite{pmfg05}, \cite{asset07}, and \cite{intro11}.

But in order to build asset graphs based on distance, we must first establish a distance measure between the nodes of the network. There are many ways to define one, but the most used in applications to financial markets is given by \cite{mst01}
\begin{equation}
\label{distance}
d_{ij}=\sqrt{2\left( 1-c_{ij}\right) }\ ,
\end{equation}
where $c_{ij}$ is the correlation between nodes $i$ and $j$. As correlations between indices vary from $-1$ (anticorrelated) to $1$ (completely correlated), the distance between them vary from $0$ (totally correlated) to $2$ (completely anticorrelated). Totally uncorrelated indices would have distance $1$ between them.

Based on the distance measures, two-dimensional coordinates are assigned to each index using an algorithm called Classical Multidimensional Scaling \cite{borg}, which is based on minimizing the stress function
\begin{equation}
\label{stress}
S=\left[ \frac{\displaystyle{\sum_{i=1}^n\sum_{j>i}^n\left( \delta_{ij}-\bar d_{ij}\right) ^2}}{\displaystyle{\sum_{i=1}^n\sum_{j>i}^nd_{ij}^2}}\right] ^{1/2}\ \ ,\ \ \bar d_{ij}=\left[ \sum_{a=1}^m\left( x_{ia}-x_{ja}\right) ^2\right] ^{1/2}\ .
\end{equation}
where $\delta_{ij}$ is 1 for $i=j$ and zero otherwise, $n$ is the number of rows of the correlation matrix, and $\bar d_{ij}$ is an m-dimensional Euclidean distance (which may be another type of distance for other types of multidimensional scaling).

The outputs of this optimization problem are the coordinates $x_{ij}$ of each of the nodes, where $i=1,\cdots, n$ is the number of nodes and $j=1,\cdots ,m$, being $m$ the number of dimensions in an $m$-dimensional space. The true distances are only perfectly representable in $m=n$ dimensions, but it is possible for a network to be well represented in smaller dimensions. In the case of this article we shall consider $m=2$ for a 2-dimensional visualization of the network. The choice is a compromise between fidelity to the original distances and the easiness of representing the networks.

In Figure 17 (left graph), we plot all indices on their coordinates. The indices are represented by the four first letters of each country, with the exception of Austria, which is written {\sl Autr} so as to differentiate it from Australia, and by three or fours letters in case of countries with multiple words, like CzRe for the Czech Republic or USA for the United States of America. Lagged indices are represented by black boxes and the original ones are represented in white boxes. Note that there is a clear distinction between original and lagged indices, forming two separate clusters with some interaction occurring for indices which are not very correlated to any other indices in the network, like some Arab and African indices. There is also a clear clustering of indices according to geography (or by time zones), particularly for Western European indices, American, and Pacific Asian ones.

By using a false color map of the expanded correlation matrix, right graph in Figure 17, one can also perceive the same organization seen in the network of indices: two main blocks, with original and lagged indices, and internal blocks according to geographic regions. Since the correlation matrix is symmetric, only one of the triangles, on the top or bellow the main diagonal, are of relevance. On the lower left quarter of the matrix, we have the correlations of the original indices with themselves. The correlations between the lagged and the original indices are part of the lower right quarter of the correlation matrix, and show little correlation between original and lagged indices, although some correlation of lagged American and European indices with the ones of Pacific Asia may be seen. There is also a brighter region on the diagonal of this quarter of the correlation matrix, showing very feeble correlations of the indices with their next day counterparts.

As some correlations may be the result of random noise, we ran some simulations based on randomized data, consisting on randomly reordering the time series of each index so as to destroy any true correlations between them but maintain each frequency distribution intact. The result of 1,000 simulations is a distance value above which correlations are probably due to random noise, which, in our case, is $d=1.36$.

The asset graphs are built using distance measures as thresholds. As an example, for threshold $T=0.5$, one builds an asset graph where all distances bellow $0.5$ are represented as edges (connections) between nodes (indices). All distances above this threshold are removed, and all indices that do not connect to any other index bellow this threshold are also removed.

Figure 18 shows the asset graphs for the world stock market indices, displaying only the indices that are connected bellow certain distance thresholds ($T=0.5$, $T=0.7$, $T=0.9$, and $T=1.1$) and their connections. For $T=0.5$, the only countries that are connected are France, Germany, the Netherlands, and Italy, original and lagged ones, forming the core of the Central European countries. For $T=0.7$, the UK, Switzerland, Belgium, Sweden, Finland, and Spain join the European cluster, which is represented in its two versions (original and lagged).

At $T=0.9$, two more clusters are formed: one of American indices, made by the USA, Canada, Mexico, and Brazil, and one of Pacific Asian indices, made by Japan, Hong Kong, South Korea, Singapore, and Taiwan. The European cluster grwos, with the addition of some more European countries. For $T=1.1$, the clusters join and add to themselves many more indices. Something to be noticed is that Israel and South Africa are part of the European cluster, and that Australia and New Zealand join the Pacific Asian cluster. Most important, there are now connections between lagged and original indices, with the lagged indices of the USA, Canada, Mexico, and Brazil forming connections with the next day indices of Japan, Hong Kong, the Philippines, Australia, and New Zealand.

The same type of asset graphs can be built for other periods of time, without much change. The network structure persists and is robust in time, and different thresholds may be used in order to analyze clusters formed at different strengths of connections \cite{Leocluster}. The use of an enlarged correlation matrix with the lagged indices by one and by two days, with $3\times 97=291$ indices, shows a similar structure (not shown here) of West to East to West influence.

What we may conclude from both the network structure and from the false color map of the expanded correlation matrix is that stock market indices are much more correlated among themselves, and by geographical or time-zone region than with their lagged values by one day. The exceptions are some indices that are very weakly correlated with the general network. We also detect some correlation of American indices with some indices from Australasia, indicating some sort of influence of the former on the latter.

The structure of the network of stock market indices may be used in order to study the propagation of crises. There is a vast literature on contagion between banks or other markets via networks \cite{Contagion01}-\cite{Contagion43}. In the majority of these works, the interbank loans and liabilities are used in order to build networks of financial institutions. A general result is that contagion may occur for a certain critical level of integration between institutions, and that high integration may be both a curse and a blessing for the propagation of crises. This will not be discussed in this article, but we stress that certain measures of centrality may be useful in the identification of the main agents in the propagation of a crisis.

Some of the main centrality measures used in the theory of Complex Networks are {\sl node centrality}, which measures the number of connections a node has, {\sl eigenvector centrality}, which measures how connected a node is, and how connected its neighbors are, and {\sl betweenness centrality}, which measures if a node is in the shortest path connecting other nodes in the network \cite{Newman}. The application of these measures of centrality to asset graphs gives results that depend on the threshold level that is chosen \cite{Leocluster}. As an example, for low enough values of the threshold, there are no connections between nodes, and for high enough threshold values, all nodes are connected to every other node in the network.

If these measures are applied to the asset graph for $T=1.1$, we obtain the results listed in Table 2, which contains the nodes with the ten best scores in each of the centrality measures. Central European indices occupy the first positions both for node centrality and eigenvector centrality, since they are very connected and in a region of highly connected nodes. So, Europe would be a center for the propagation of crises in a variety of models. Now for betweenness centrality, we have Singapore and Japan occupying the first two positions, mainly due to their role as connections between Pacific Asian and European indices and between the lagged indices of the Americas and the indices of Eurasia. The USA also occupies a high rank due to the same reason.

\vskip 0.3 cm

{\footnotesize
\[ \begin{array}{|c|c|c|} \hline \text{Node Degree} & \text{Eigenvector Centrality} & \text{Betweenness Centrality} \\ \hline \text{France} & \text{UK, France} & \text{Singapore} \\ \text{UK} & \text{Netherlands} & \text{Japan} \\ \text{Netherlands} & \text{Belgium, Italy, Germany} & \text{USA} \\ \text{Finland, Italy, Belgium, Germany, Norway} & \text{Finland} & \text{France} \\ \text{Sweden, Spain} & \text{Sweden, Spain} & \text{Australia} \\ \text{Australia} & \text{Norway} & \text{Netherlands} \\ \text{Switzerland} & \text{Austria} & \text{Finland} \\ \text{Denmark, Czech Republic, USA} & \text{Switzerland} & \text{Hong Kong} \\ \text{Poland} & \text{Czech Republic} & \text{Austria} \\ \text{South Africa, Hungary, Ireland} & \text{Denmark, Poland} & \text{Denmark} \\ \hline \end{array} \]
}

\noindent {\bf Table 2.} Highest ranking stock market indices according to node centrality, eigenvector centrality, and betweenness centrality. Only the ten most central classifications are shown.

\vskip 0.3 cm

\section{Conclusions}

We analyzed from various aspects the issue of lagging or not international stock market indices when calculating their correlations. We established that indices of stock exchanges that have intersecting hours with the New York Stock Exchange have larger correlations with the S\&P 500 than their lagged counterparts, and that indices that have non-intersecting hours with the same stock exchange have larger correlations when their indices are compared with the S\&P 500 of the previous day, replicating the results of previous authors. The frequency distributions of the eigenvalues of the correlation matrices between international stock market indices and the eigenvectors associated with the largest eigenvalues were also analyzed, showing the presence of a structure that is typical of stock markets that operate at different times when looking at the eigenvectors associated with the second largest eigenvalues of the correlation matrices.

A solution for the issue of lagging or not indices was proposed, in which, if one wants to study the relations and dynamics of the world stock market indices, one must then consider the cyclic chain of influences, and probably decide to adopt a model where there are both original and lagged indices. This solution was implemented and tested from the points of view of both the eigenvalues and eigenvectors of the resulting enlarged correlation matrix and of a network structure derived from the enlarged correlation matrix. Both show a structure consisting of two blocks, of original and lagged indices, with interactions between the lagged indices of some American stock exchanges with indices of stock exchanges of countries from the Eurasia. A study of the centrality of the indices when seen on a network built as an asset graph shows that Central European indices occupy a position of stronger centrality and form a block of highly central indices, and that Singapore, Japan, and the USA are the main vehicles between interactions because of their positions as intermediators between lagged and original indices.

This work may be used in further studies of causality relations between stock markets around the globe and on a better understanding of how crises propagate across markets. Its main contribution is the proposal of considering both original and lagged indices when working with stock markets or, indeed, any markets, that operate at different times.

\vskip 0.6 cm

\noindent{\bf Acknowledgements}

\vskip 0.4 cm

The author acknowledges the support of this work by a grant from Insper, Instituto de Ensino e Pesquisa. This article was written using \LaTeX, all figures were made using PSTricks and Matlab, and the calculations were made using Matlab and Excel. All data are freely available upon request on leonidassj@insper.edu.br.

\begin{figure}[H]
\begin{center}
\includegraphics[scale=0.8]{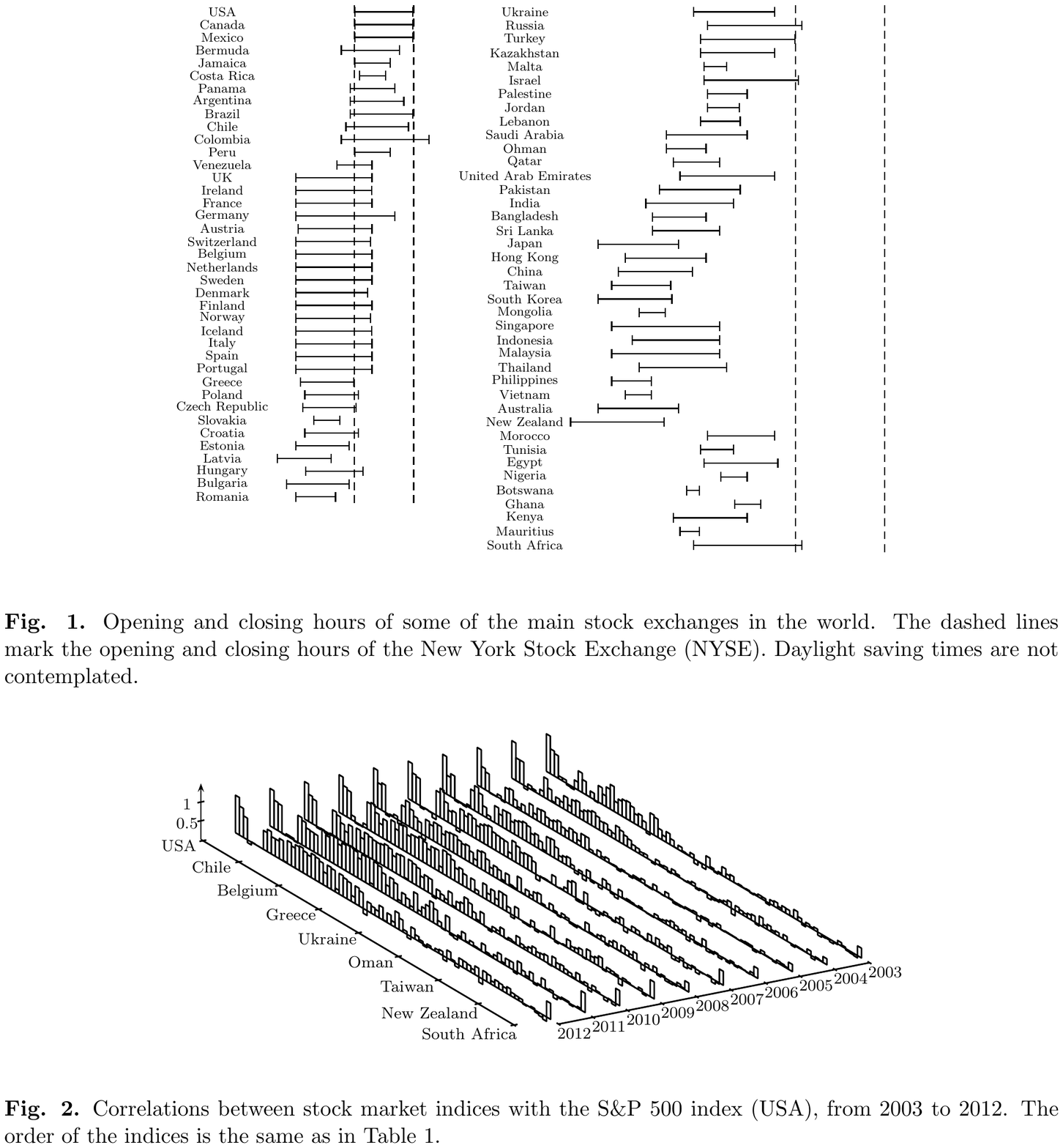}
\end{center}
\end{figure}

\begin{figure}[H]
\begin{center}
\includegraphics[scale=0.8]{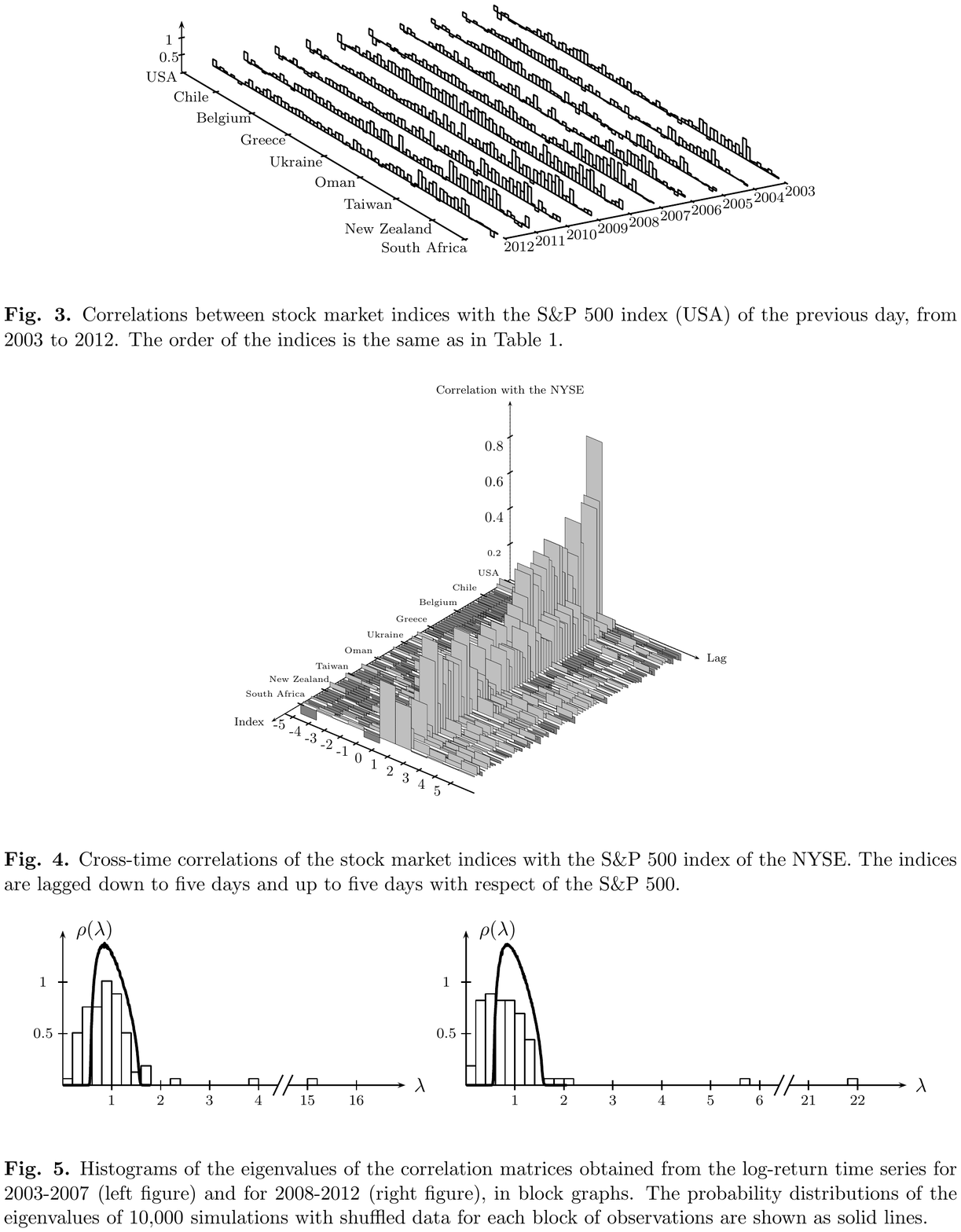}
\end{center}
\end{figure}

\begin{figure}[H]
\begin{center}
\includegraphics[scale=0.8]{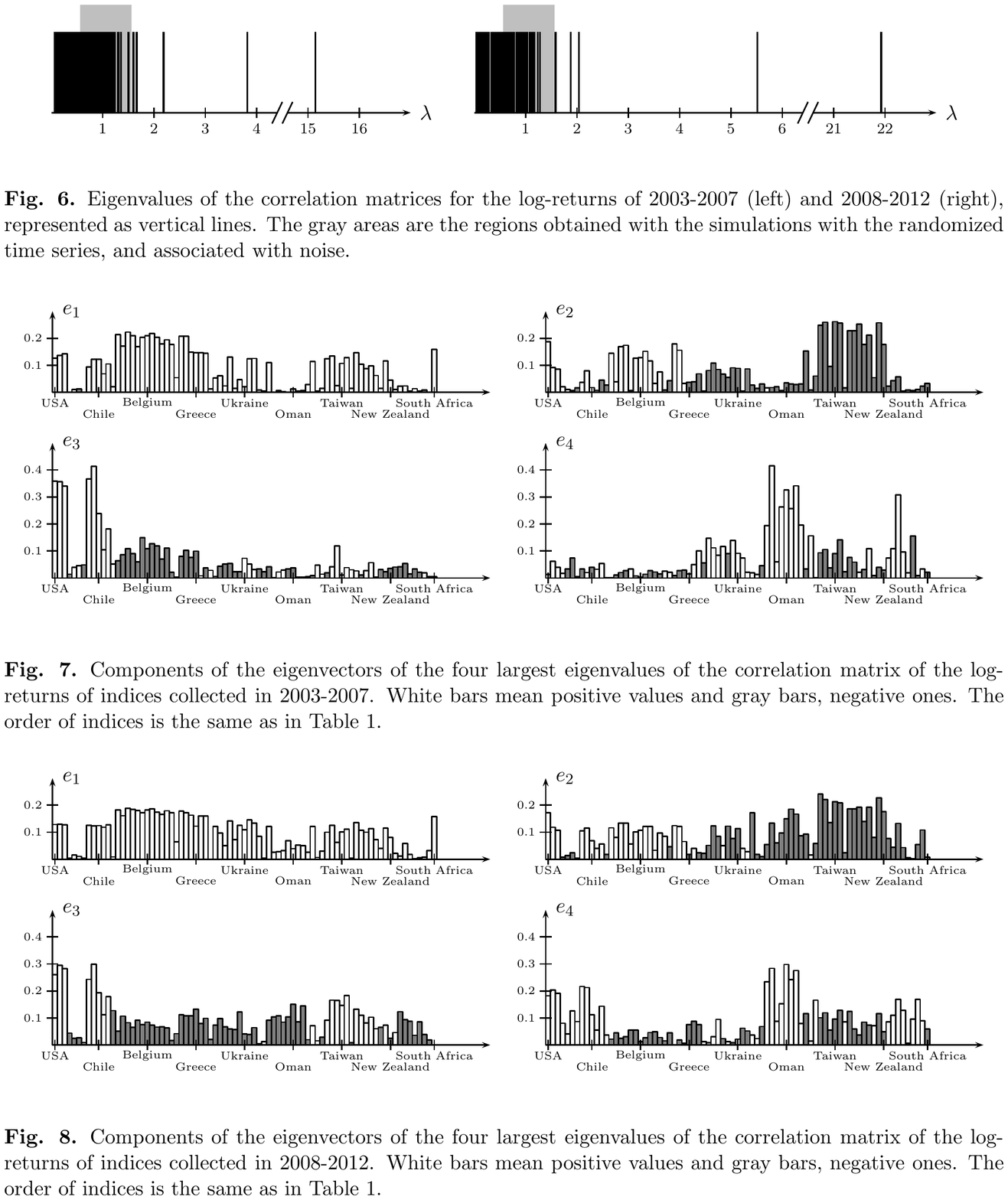}
\end{center}
\end{figure}

\begin{figure}[H]
\begin{center}
\includegraphics[scale=0.8]{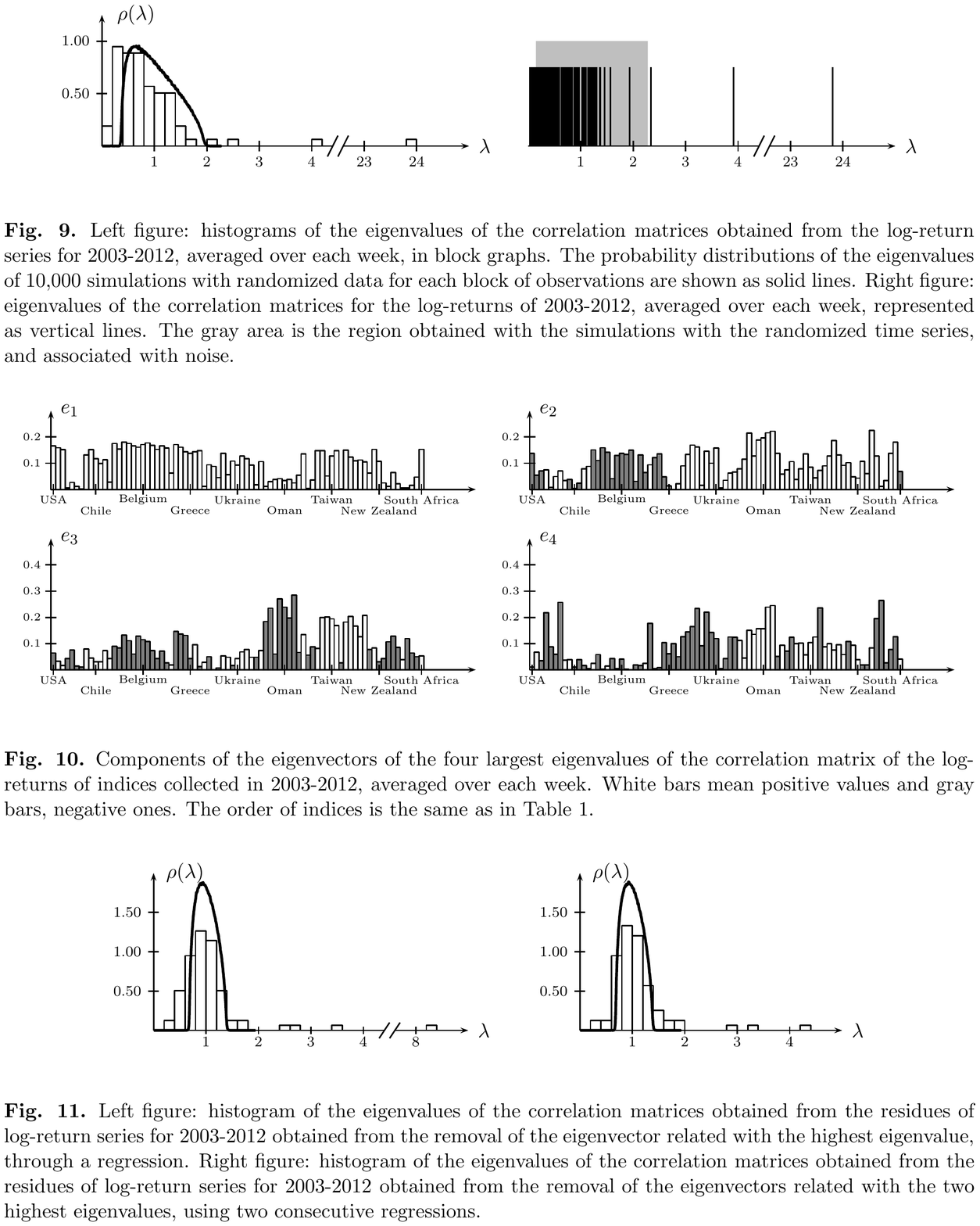}
\end{center}
\end{figure}

\begin{figure}[H]
\begin{center}
\includegraphics[scale=0.8]{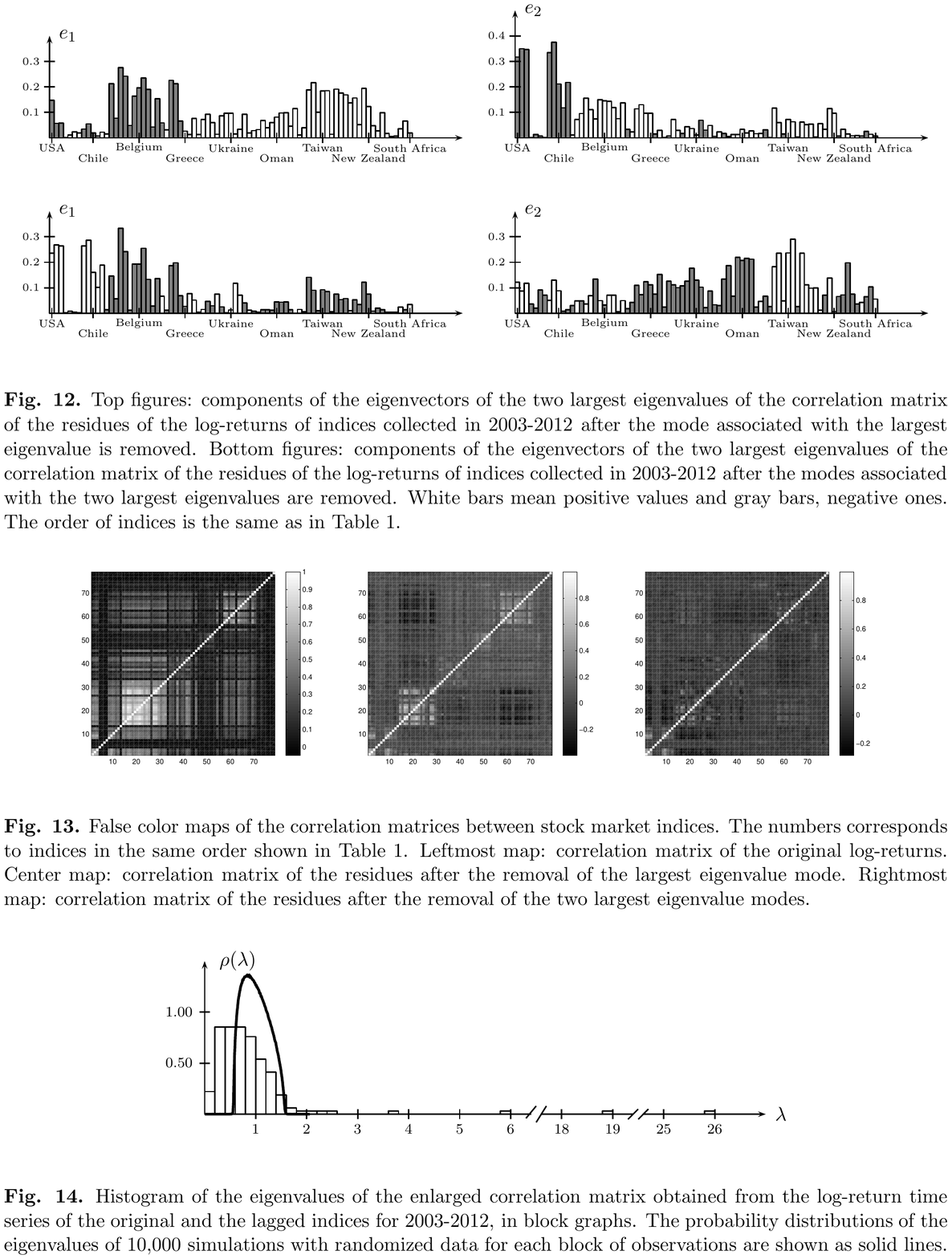}
\end{center}
\end{figure}

\begin{figure}[H]
\begin{center}
\includegraphics[scale=0.8]{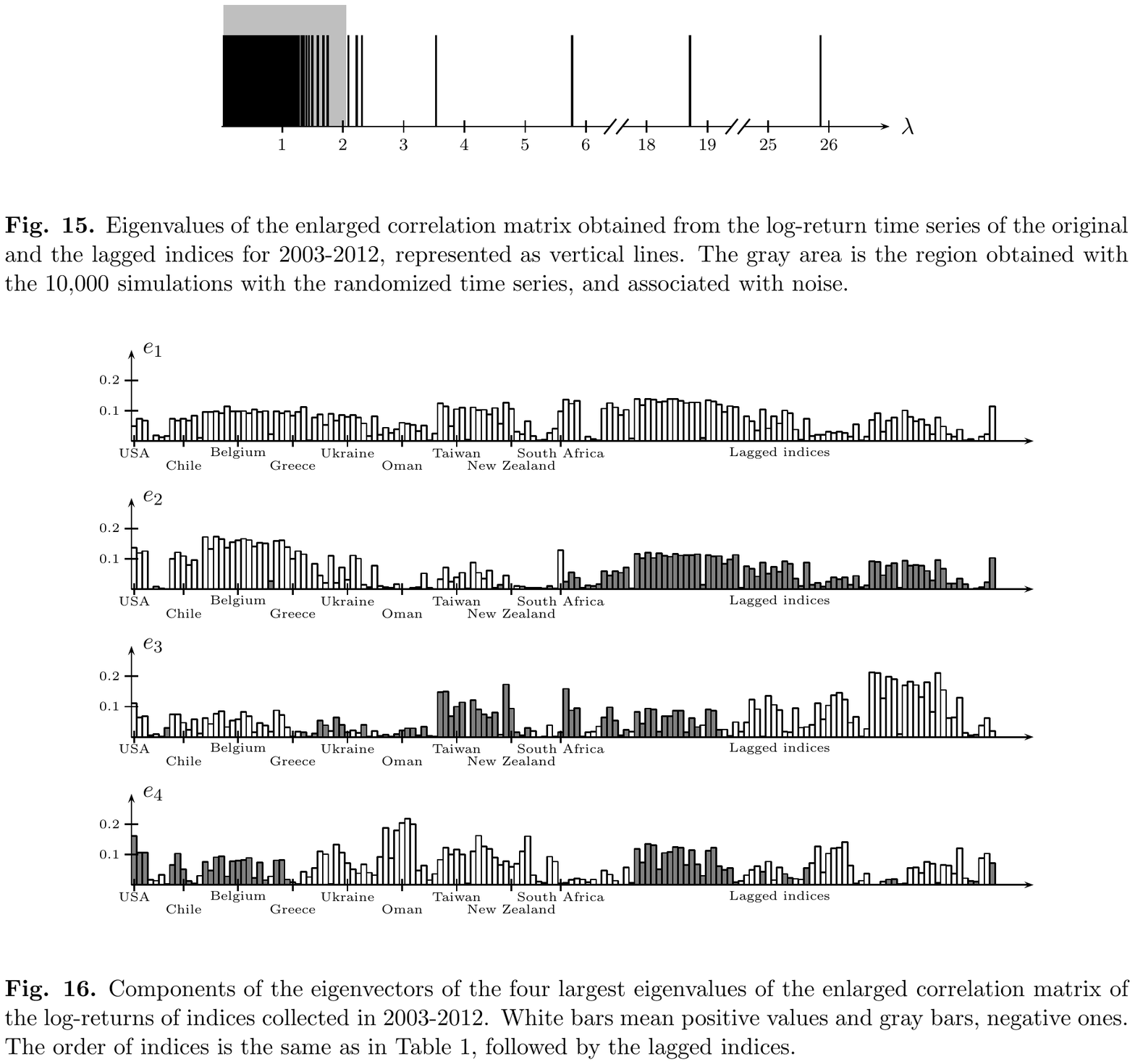}
\end{center}
\end{figure}

\begin{figure}[H]
\begin{center}
\includegraphics[scale=0.8]{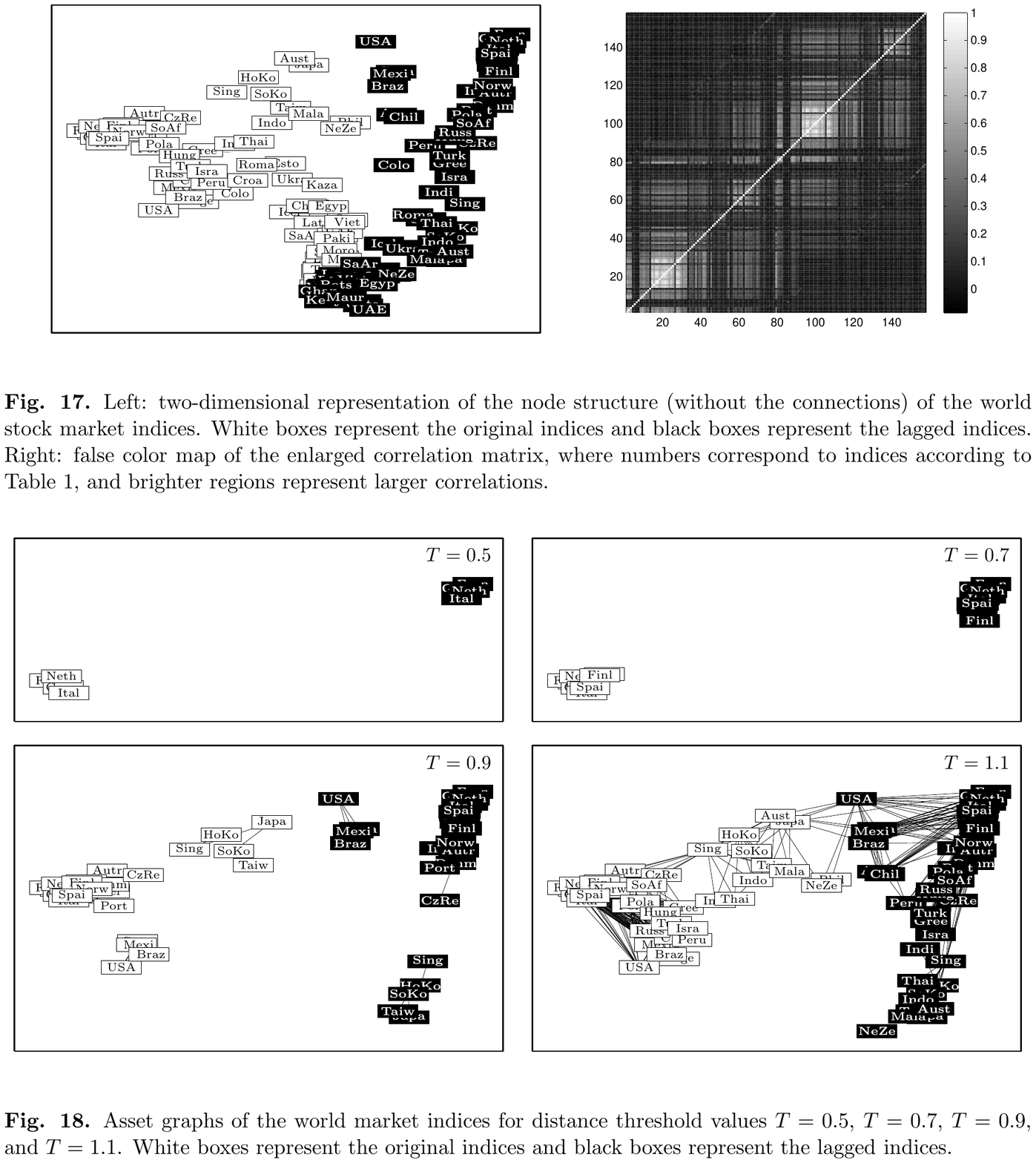}
\end{center}
\end{figure}


\vskip 0.5 cm

\begin{thebibliography}{99}


\bibitem{ES1989} C.S. Eum and S. Shim, {\sl International transmission of stock market movements}, The Journal of Financial and Quantitative Analysis 24 (1989) 241-256.

\bibitem{BFG1990} K.G. Becker, J.E. Finnerty, and M. Gupta, {\sl The intertemporal relation between the U.S. and Japanese stock markets}, The Journal of Finance 45 (1990) 1297-1306.

\bibitem{LEI1994} W.L. Lin, R.F. Engle, and T. Ito, {\sl Do bulls and bears move across borders? International transmission of stock returns and volatility}, Review of Financial Studies 7 (1994), 507-538.

\bibitem{B1996} T.J. Brailsford, {\sl Volatility spillovers across the Tasman}, Australian Journal of Management 21 (1996) 13-27.

\bibitem{BVM2000} G. Bonanno, N. Vanderwalle, and R.N. Mantegna, {\sl Taxonomy of stock market indices}, Phys. Rev. E 62 (2000), R7615-R7618.

\bibitem{MA2000} G.M. Mian and C.M. Adam, {\sl Does more market-wide information originate while an exchange is open: some anomalous evidence from the ASX}, Australian Journal of Management 25 (2000) 339-352.

\bibitem{VBB2000} N.Vandewalle, Ph.Boveroux and F.Brisbois, {\sl Domino effect for world market fluctuations}, The European Physics Journal B 15 (2000) 547-549.

\bibitem{DGRS2001} Dro\.{z}d\.{z}, Grümmer, Ruf, and Speth , {\sl Towards identifying the world stock market cross-correlations: DAX versus Dow Jones}, Physica A 294 (2001) 226-234.

\bibitem{M2001} S. Maslov, {\sl Measures of globalization based on cross-correlations of world financial indices}, Physica A {\bf 301} (2001) 397-406.

\bibitem{MA2006} K. B. K. Mayya and R. E. Amritkar , {\sl Analysis of delay correlation matrices}, Proceedings of the Third National Conference on Nonlinear Systems and Dynamics (2006) p. 183. E-print cond-mat/0601279.

\bibitem{KDGO2006} J. Kwapie\'{n}, S. Dro\.{z}d\.{z}, A.Z. Górski, and P. O\'swi\c{e}cimka, {\sl Asymmetric matrices in an analysis of financial correlations}, Acta Physica Polonica B 37 (2006) 3039-3048.

\bibitem{STZM2011} D-M Song, M. Tumminello, W-X Zhou, and R.N. Mantegna, {\sl Evolution of worldwide stock markets, correlation structure, and correlation-based graphs}, Physical Review E 84 (2011) 026108.

\bibitem{LR2012} G. Livan and L. Rebecchi, {\sl Asymmetric correlation matrices: an analysis of financial data}, The European Physics Journal B 85 (2012) 213.

\bibitem{NMHL2013} A. Nobi, S.E. Maeng, G.G. Ha, and J.W. Lee, {\sl Random Matrix Theory on cross-correlations in global financial indices and local stock market indices}, (2013) arXiv:1302.6305v1.

\bibitem{L2013} F. Li, {\sl Identifying asymmetric comovements of international stock market returns}, Journal of Financial Econometrics (online), first published online in June 7, 2013.


\bibitem{LM1990} A. Lo and A.C. MacKinlay, {\sl An econometric analysis of nonsynchronous trading}, Journal of Econometrics, Elsevier 45 (1990) 181-211.

\bibitem{JN1997} F. de Jong and T. Nijman, {\sl High frequency analysis of lead-lag relationships between financial markets}, Journal of Empirical Finance 4 (1997) 259-277.

\bibitem{MSG2010} M.C. Münnix, R. Schäfer, and T. Guhr, {\sl Compensating asynchrony effects in the calculation of financial correlations}, Physica A 389 (2010) 767-779.

\bibitem{ZY2012} Z. Zheng and K. Yamasaki, {\sl Hierarchical structure and time-lag correlation in worldwide financial markets}, arXiv:1205.1861v1 (2012).

\bibitem{HA2012} N. Huth and F. Abergel, {\sl High frequency lead/lag relationships - empirical facts}, arXiv:1111.7103.

\bibitem{HY2005} T. Hayashi and N. Yashida, {\sl On covariance estimation of non-synchronously observed diffusion processes}, Bernoulli 11 (2005) 359-379.

\bibitem{GAMK2012} B. Goswamia, G. Ambikaa, N. Marwanb, and J. Kurths, {\sl On interrelations of recurrences and connectivity trends between stock indices}, (2011) arXiv:1103.5189v1.

\bibitem{CPR1} N. Marwan, M.C. Romano, M. Thiel, and J. Kurths, {\sl Recurrence plots for the analysis of complex systems}, Physics Reports 438 (2007) 237–329.

\bibitem{CPR2} M.C. Romano, M. Thiel, J. Kurths, I.Z. Kiss, and J.L. Hudson, {\sl Detection of synchronization for non-phase-coherent and non-stationary data}, Europhysics Letters 71 (2005) 466–472.

\bibitem{BC2010} J.A. Bastos and J. Caiado, {\sl Recurrence quantification analysis of global stock markets}, (2010) in http://cibs.ck.ua/files/scipub/piskun01.pdf.

\bibitem{KD2012} S. Kumar and N. Deo, {\sl Correlation, network and multifractal analysis of global financial indices}, (2012) arXiv:1202.0409v1.

\bibitem{rmt1} M.L. Mehta, {\sl Random Matrices}, (2004) Academic Press.

\bibitem{mfdfa} J.W. Kantelhardt, S.A. Zschiegner, E. Koscielny-Bunde, S. Havlin, A. Bunde, and H.E. Stanley, {\sl Multifractal detrended fluctuation analysis of nonstationary time series}, Physica A 316 (2002) 87-114.


\bibitem{Leocorr} L. Sandoval Jr. and I. De P. Franca, {\sl Correlation of financial markets in times of crisis}, Physica A 391 (2012) 187-208.

\bibitem{Leotree} L. Sandoval Jr., {\sl Pruning a minimum spanning tree}, Physica A 391 (2012) 2678-2711.

\bibitem{Leocluster} L. Sandoval Jr., {\sl Cluster formation and evolution in networks of financial market indices}, Algorithmic Finance 2 (2013) 3-43.


\bibitem{pruning} L. Sandoval Jr., {\sl Pruning a Minimum Spanning Tree}, Physica A 391 (2012) 2678–2711.

\bibitem{rmt3} Mar\v{e}nko, V.A.; Pastur, L.A. USSR-Sb { 1} (1967) 457-483.

\bibitem{LeoBovespa} L. Sandoval Jr., A. Bruscato, and M.K. Venezuela, {\sl Building portfolios of stocks in the São Paulo Stock Exchange using Random Matrix Theory}, arXiv:1201.0625.


\bibitem{mst01} R.N. Mantegna, {\sl Hierarchical structure in financial markets}, The European Phys. J. B {\bf 11} (1999) 193.

\bibitem{mst02} G. Bonanno, F. Lillo, and R.N. Mantegna, {\sl High-frequency cross-correlation in a set of stocks}, Quantitative Finance {\bf 1} (2001) 96-104.

\bibitem{mst03} S. Micchichè, G. Bonanno, F. Lillo, and R.N. Mantegna, {\sl Degree stability of a minimum spanning tree of price return and volatility}, Physica A {\bf 324} (2003) 66-73.

\bibitem{mst04} G. Bonanno, G. Caldarelli, F. Lillo, S. Miccichè, N. Vandewalle, and R.N. Mantegna, {\sl Networks of equities in financial markets}, The European Phys. J. B {\bf 38}, (2004) 363–371.

\bibitem{mst05} R. Coelho, C.G. Gilmore, B. Lucey, P. Richmond, and S. Hutzler, {\sl The evolution of interdependence in world equity markets - evidence from minimum spanning trees}, Physica A {\bf 376} (2007) 455-466.

\bibitem{mst06} C. Borghesi, M. Marsili, and S. Miccichè, {\sl Emergence of time-horizon invariant correlation structure in financial returns by subtraction of the market mode}, Phys. Rev. E {\bf 76} (2007), 026104.

\bibitem{mst07} J.G. Brida and W.A. Risso, {\sl Multidimensional minimal spanning tree: the Dow Jones case}, Physica A {\bf 387} (2008) 5205-5210.

\bibitem{mst09} J. Kwapie\'{n}, S. Gworek, S. Dro\.{z}d\.{z}, and A. Górski, {\sl Analysis of a network structure of the foreign currency exchange market}, J. of Economic Interaction and Coordination {\bf 4} (2009), 55-72.

\bibitem{mst10} J. Kwapie\'{n}, S. Gworek, and S. Dro\.{z}d\.{z}, {\sl Structure and evolution of the foreign exchange networks}, Acta Physica Polonica B {\bf 40} (2009), 175-194.

\bibitem{mst11} S. Dro\.{z}d\.{z}, J. Kwapie\'{n}, and J. Speth, {\sl Coherent patterns in nuclei and in financial markets}, AIP Conf. Proc. {\bf 1261} (2010), 256-264.

\bibitem{mst13} Y. Zhanga, G.H.T. Leea, J.C. Wonga, J.L. Kokb, M. Prustyb, and S.A. Cheong, {\sl Will the US Economy Recover in 2010? A Minimal Spanning Tree Study}, (2010) Physica A {\bf 390} (2011) 2020-2050.

\bibitem{mst14} M. Keskin, B. Deviren, and Y. Kocalkaplan, {\sl Topology of the correlation networks among major currencies using hierarchical structure methods}, Physica A {\bf 390} (2011), 719-730.


\bibitem{pmfg01} M. Tumminello, T. Aste, T. Di Matteo, R.N. Mantegna, {\sl A tool for filtering information in complex
systems}, Proceedings of the National Academy of Sciences of the United States of America {\bf 102}, (2005) 10421-10426.

\bibitem{pmfg02} C. Coronnello, M. Tumminello, F. Lillo, S. Micchichè, and R.N. Mantegna, {\sl Sector identification in a set of stock return time series traded at the London Stock Exchange}, Acta Phys. Pol. B {\bf 36} (2005) 2653-2679.

\bibitem{pmfg03} T. Aste and T. Di Matteo, {\sl Correlation filtering in financial time series}, {\sl Noise and Fluctuations in Econophysics and Finance}, Proceedings of the SPIE 5848 (2005) 100-109.

\bibitem{pmfg04} M. Tumminello, T. Di Matteo, T. Aste, and R.N. Mantegna, {\sl Correlation based networks of equity returns sampled at different time horizons}, Eur. Phys. J. B {\bf 55} (2007) 209–217.

\bibitem{pmfg05} C. Coronnello, M. Tumminello, F. Lillo, S. Micchichè, and R.N. Mantegna, {\sl Economic sector identification in a set of stocks traded at the New York Stock Exchange: a comparative analysis}, Proceedings of the SPIE, vol. 6601, 66010T (2007).

\bibitem{pmfg07} M. Tumminello, F. Lillo, and R.N. Mantegna, {\sl Correlation, hierarchies, and networks in financial markets}, Journal of Economic Behavior \& Organizations {\bf 75} (2010), 40-58.

\bibitem{pmfg08} D.Y. Kenett, M. Tumminello, A. Madi, G. Gur-Gershgoren, R.N. Mantegna, and E. Ben-Jacob, {\sl Dominating Clasp of the Financial Sector Revealed by Partial Correlation Analysis of the Stock Market}, PLoS ONE 5 (12) (2010) e15032.


\bibitem{asset01} J.-P. Onnela, A. Chakraborti, K. Kaski, and J. Kertész, {\sl Dynamic asset trees and portfolio analysis}, The European Phys. J. B {\bf 30} (2002) 285-288.

\bibitem{asset02} J.-P. Onnela, A. Chakraborti, and K. Kaski, {\sl Dynamics of market correlations: taxonomy and portfolio analysis}, Phys. Rev. E {\bf 68} (2003) 1-12.

\bibitem{asset03} J.-P. Onnela, A. Chakraborti, K. Kaski, and J. Kertész, {\sl Dynamic asset trees and Black Monday}, Physica A {\bf 324} (2003) 247-252.

\bibitem{asset04} J.-P. Onnela, A. Chakraborti, K. Kaski, J. Kertész, and A. Kanto, {\sl Asset trees and asset graphs in financial markets}, Phys. Scripta T {\bf 106} (2003) 48-54.

\bibitem{asset05} J.-P. Onnela, K. Kaski, and J. Kertész, {\sl Clustering and information in correlation based financial networks}, Eur. Phys. J. B {\bf 38} (2004) 353-362.

\bibitem{asset06} S. Sinha and R.K. Pan, {\sl Uncovering the internal structure of the Indian financial market: cross-correlation behavior in the NSE}, in ``Econophysics of markets and business networks'', Springer (2007) 215-226.

\bibitem{asset07} M. Ausloos and R. Lambiotte, {\sl Clusters or networks of economies? A macroeconomy study through gross domestic product}, Physica A {\bf 382} (2007) 16-21.

\bibitem{asset08} L. Sandoval Jr., {\sl A Map of the Brazilian Stock Market}, Advances in Complex Systems {\bf 15} (2012) 1250042-1250082.


\bibitem{intro11} M. Eryi\v{g}it and R. Eryi\v{g}it, {\sl Network structure of cross-correlations among the world market indices}, Physica A {\bf 388} (2009), 3551-3562.


\bibitem{borg} I. Borg and P. Groenen, {\sl Modern Multidimensional Scaling: theory and applications}, 2nd edition, (2005) Springer-Verlag.


\bibitem{Contagion01} Kirman, A. {\sl The economy as an evolving network}, Journal of Evolutionary Economics { 7} (1997) 339-353.

\bibitem{Contagion02} Allen, F.; Gale, D. {\sl Financial contagion}, Journal of Political Economy { 108} (2000) 1-33.

\bibitem{Contagion03} Watts, D. {\sl A simple model of global cascades on random networks}, Proceedings of the National Academy Sciences { 99} (2002) 5,766-5,771.

\bibitem{Contagion04} Nagurney, A. {\sl Financial and economic networks: an overview}, in Innovations in financial and economic networks (2003), Edward Elgar Publishibg, 1-25.

\bibitem{Contagion05} Boss, M.; Elsinger, H.; Thurner, S.; Summer, M. {\sl An empirical analysis of the network structure of the Austrian interbank market}, Österreichischer Nationalbank Financial Stability Report (2004) 77-87.

\bibitem{Contagion06} Vivier-Lirimont, S. {\sl Interbanking networks: towards a small financial world?}, Cahiers de la Maison des Sciences Economiques v04046 (2004), Université Panthon-Sorbonne (Paris 1).

\bibitem{Contagion07} Eboli, M. {\sl Systemic risk in financial networks: a graph theoretic approach}, Universitá di Chieti Pescara (2004).

\bibitem{Contagion08} Boss, M.; Elsinger, H.; Thurner, S.; Summer, M. {\sl The network topology of the interbank market}, Quantitative Finance { 4} (2004) 677-684.

\bibitem{Contagion09} Leitner, Y. {\sl Financial networks: contagion, commitment and private sector bailouts}, Journal of Finance { 60} (2005) 2,925-2,953.

\bibitem{Contagion10} Iori, G.; Masi, G.; Precup, O.V.; Gabbi, G.; Caldarelli, G. {\sl A network analysis of the Italian overnight money market}, Discussion Paper Series 05/05, City University London (2005).

\bibitem{Contagion11} Leitner, Y. {\sl Financial networks: contagion, commitment and private sector bailouts}, Journal of Finance { 60} (2005) 2925-2953.

\bibitem{Contagion12} Markose, S.M. {\sl Computability and evolutionary complexity: markets as complex adpative systems (CAS)}, Economic Journal { 115} (2005) F159-F192.

\bibitem{Contagion13} Muller, J. {\sl Interbank credit lines as a channel of contagion}, Journal of Financial Services Research { 29} (2006) 37-60.

\bibitem{Contagion14} Soramäki, K.; Bech, M.L.; Arnold, J.; Glass, R.J.; Beyeler, W.E. {\sl The topology of interbank payment flows}, Physica A { 379} (2007) 317333.

\bibitem{Contagion15} Nier, E.; Yang, J.; Yorulmazer, T.; Alentorn, A. {\sl Network models and financial stability}, Journal of Economic Dynamics and Control (2007).

\bibitem{Contagion16} Kleinberg, J. {\sl Cascading behavior in networks: algorithmic and economic issuses}, in Algorithmic Game Theory, 613-632, Cambridge University Press (2007).

\bibitem{Contagion17} Castiglionesi, F.; Navarro, N. {\sl Optimal fragile financial networks}, Tilburg University Discussion Paper no. 2007-100 (2007).

\bibitem{Contagion18} Nier, E.; Yang, J.; Yorulmazer, T.; Alentorn, A. {\sl Network models and financial stability}, Journal of Economic Dynamics and Control { 31} (2007) 2,033-2060.

\bibitem{Contagion19} Soramali, K.; Bech, M.L.; Arnold, J.; Glass, R.J.; Beyeler, W.E. {\sl The topology of interbank payment flows}, Physica A { 379} (2007) 317-333.

\bibitem{Contagion20} Cossin, D.; Schellhorn, H. {\sl Credit risk in a network economy}, Management Science { 53} (2007) 1,604-1,617.

\bibitem{Contagion21} Lorenz, J.; Battiston, S.; Schweitzer, F. {\sl Systemic risk in a unifying framework for cascading processes on networks}, European Physics Journal B { 71} (2009) 441-460.

\bibitem{Contagion22} Haldane, A. {\sl Rethinking the financial network}, speech delivered at the Financial Student Association, Amsterdam, Aprol (2009).

\bibitem{Contagion23} Allen, F.; Babus, A. {\sl Networks in finance}, in Kleindorfer, P., Wing, Y.,and Gunther, R. (eds.), {\sl The network challenge: strategy, profit, and risk in an interlinked world}, (2009) Wharton Scool Publishing.

\bibitem{Contagion24} Chan-Lau, J.; Espinosa, M.; Solé, J. {\sl On the use of network analysis to assess systemic financial linkages}, (2009) IMF Global Financial Stability Report, Vol. 2, April 2009. Available at SSRN: http://ssrn.com/abstract=1417920.

\bibitem{Contagion25} Georg, C-P; Poschmann, J. {\sl Sistemic risk in a network model of interbank markets with central bank activity}, Jena Economic Research Papers (2010), University of Jena.

\bibitem{Contagion26} Cont, R.; Moussa, A.; Santos, E.B. {\sl Network Structure and Systemic Risk in Banking Systems} (2010). Available at SSRN: http://ssrn.com/abstract=1733528 or http://dx.doi.org/10.2139/ssrn.1733528

\bibitem{Contagion27} Gai, P.; Kapadia, S. {\sl Contagion in Financial Networks}, Proceedings of the Royal Society A { 466} (2010).

\bibitem{Contagion28} Canedo, J.M.D.; Martínez-Jaramillo, S. {\sl Financial contagion: a network model for estimating the distribution of loss for the financial system}, Journal of Economic Dynamics and Control { 34} (2010) 2358-2374.

\bibitem{Contagion29} Cohen-Cole, E.; Patacchini, E.; Zenou, Y. {\sl Systemic risk and network formation in the interbank market}, (2010) CAREFIN Research Paper No. 25/2010, University of Stockholm. Available at SSRN: http://ssrn.com/abstract=1799925.

\bibitem{Contagion30} Dette, T.; Pauls, S.; Rockmore, D.N. {\sl Robustness and contagion in the international financial network} (2010) arXiv:1104.4249v2.

\bibitem{Contagion31} Georg, C-P {\sl The Effect of the Interbank Network Structure on Contagion and Financial Stability}, Working Papers on Global Financial Markets, No. 12 (2010).

\bibitem{Contagion32} Markose, S.; Giansante, S.; Gatkowski, M.; Shaghaghi, A.R. {\sl Too Interconnected To Fail: Financial Contagion and Systemic Risk In Network Model of CDS and Other Credit Enhancement Obligations of US Banks}, COMISEF Working Papers Series WPS-033 21/04/2010 (2010).

\bibitem{Contagion33} European Central bank Report, {\sl Recent advances in modelling systemic risk using network analysis}, (2010), in http://www.ecb.int/pub/pdf/other/modellingsystemicrisk012010en.pdf.

\bibitem{Contagion34} Billio, M.; Getmansky, A.; Lo, W.; Pelizzon, L. {\sl Econometric measures of connectedness and systemic risk in the finance and insurance sectors}, (2011) Journal of Financial Economics, forthcoming.

\bibitem{Contagion35} Larry, D.; Easley, J.; Kleinberg, R.; Tardos, E. {\sl Network formation in the presence of contagious risk}, in Proceedings of the 12th ACM Conference on Electronic Commerce (2011), 1-10.

\bibitem{Contagion36} Diebold, F.X.; Yilmaz, K. {\sl On the network topology of variance decompositions: measuring the connectedness of financial firms} (2011) PIER Working Paper 11-031.

\bibitem{Contagion37} Amini, H.; Cont, R.; Minca, A. {\sl Resilience to contagion in financial networks}, (2012) arXiv:1112.5687v1.

\bibitem{Contagion38} Babus, A. {\sl The formation of financial networks}, (2012) Discussion Paper 06-093, Tinbergen Institute.

\bibitem{Contagion39} Baral, P. {\sl Strategic Behavior and endogenous risk of contagion in a financial network: a network financial game} (2012) Discussion Paper, Indiana University.

\bibitem{Contagion40} Cabrales, A.; Gottardi, P.; Vega-Redondo, F. {\sl Risk-sharing and contagion in networks} (2013) Economics Working Papers, http://EconPapers.repec.org/RePEc:cte:werepe:we1301.

\bibitem{Contagion41} Elliott, M.; Golub, B.; Jackson, M.O. {\sl Financial Networks and Contagion} (2013). Available at SSRN: http://ssrn.com/abstract=2175056.

\bibitem{Contagion42} Acemoglu, D.; Ozdaglar, A.; Tahbaz-Salehi, A. {\sl Systemic Risk and Stability in Financial Networks} (2013) NBER Working Paper No. 18727.

\bibitem{Contagion43} Glasserman, P.; Young, H.P. {\sl How likely is contagion in financial networks?} (2013) Oxford University - Department of Economics Discussion Paper Series.


\bibitem{Newman} Newman, M.E.J. {\sl Networks, and introduction}, (2010) Oxford University Press.

\end{thebibliography}
\end{document}